\documentclass[runningheads]{llncs}
\usepackage{algorithm} 
\usepackage{algorithmic}
\usepackage{booktabs}
\usepackage{clrscode3e}
\usepackage{multirow}
\usepackage{graphicx}
\usepackage{subfigure}
\usepackage{textcomp}
\usepackage{xcolor}
\usepackage{acronym}
\usepackage{amsmath}
\usepackage{amssymb}
\usepackage{amsfonts}
\usepackage{mathrsfs}
\usepackage{indentfirst}
\usepackage{cases}
\usepackage{bm}
\usepackage{color}
\usepackage{stfloats}
\usepackage{url}
\usepackage{verbatim}
\usepackage{array}

\acrodef{CSR}{Cross-domain Sequential Recommendation}
\acrodef{CDS}{Cross-domain Sequential}
\acrodef{CSRs}{Cross-domain Sequential Recommender systems}
\acrodef{GAN}{Generative Adversarial Network}
\acrodef{RNN}{Recurrent Neural Network}
\acrodef{GNN}{Graph Neural Network}
\acrodef{SRs}{Sequential Recommender Systems}
\acrodef{GCN}{Graph Convolution Network}
\acrodef{SA}{Self-Attention}
\acrodef{EA}{External Attention}
\acrodef{RNN}{Recurrent Neural Network}
\acrodef{CR}{Cross-domain Recommendation}
\acrodef{SR}{Sequential Recommendation}
\acrodef{MDP}{Markov Decision Process}
\acrodef{MLAP}{Multi-Layer Aggregating Protocol}
\acrodef{SLAP}{Single-Layer Aggregating Protocol}
\acrodef{LEA-GCN}{Lightweight External Attention-enhanced Graph Convolution Network}
\acrodef{GRU}{Gated Recurrent Unit}
\acrodef{MLP}{Multi-Layer Perceptron}

\begin{document}
\title{Towards Lightweight Cross-domain Sequential Recommendation via External Attention-enhanced Graph Convolution Network}
%
\titlerunning{Lightweight External Attention-enhanced Graph Convolution Network}
%
\author{Jinyu Zhang$^\dag$ \and Huichuan Duan$^\dag$
 \and Lei Guo$^\dag$ \and Liancheng Xu$^\dag$ \and Xinhua Wang$^\dag$\thanks{Corresponding author.}}

\authorrunning{J. Zhang et al.}
\institute{$^\dag$~School of Information Science and Engineering, Shandong Normal University\\ Jinan, China, 250358\\
\email{jinyuz1996@outlook.com,$~$leiguo.cs@gmail.com,\\$\{$hcduan,~lcxu,~wangxinhua$\}$@sdnu.edu.cn}
}
%
\maketitle 

\begin{abstract}
\ac{CSR} is an emerging yet challenging task that depicts the evolution of behavior patterns for overlapped users by modeling their interactions from multiple domains.
Existing studies on \ac{CSR} mainly focus on using composite or in-depth structures that achieve significant improvement in accuracy but bring a huge burden to the model training.
Moreover, to learn the user-specific sequence representations, existing works usually adopt the global relevance weighting strategy (e.g., self-attention mechanism), which has quadratic computational complexity.
In this work, we introduce a lightweight external attention-enhanced GCN-based framework to solve the above challenges, namely LEA-GCN. 
Specifically, by only keeping the neighborhood aggregation component and using the \acf{SLAP}, our lightweight GCN encoder performs more efficiently to capture the collaborative filtering signals of the items from both domains.
To further alleviate the framework structure and aggregate the user-specific sequential pattern, we devise a novel dual-channel \ac{EA} component, which calculates the correlation among all items via a lightweight linear structure.
Extensive experiments are conducted on two real-world datasets, demonstrating that LEA-GCN requires a smaller volume and less training time without affecting the accuracy compared with several state-of-the-art methods.

\keywords{recommendation systems \and cross-domain sequential recommendation \and attention mechanism \and graph neural network \and collaborative filtering}
\end{abstract}

\section{Introduction \label{sec:introduction}}

\noindent \ac{SR} aims to capture the user's dynamic behavioral pattern from the interaction sequence that has attracted immense research attention and has wide applications in many domains, such as electronic commerce, online retrieval, and mobile services~\cite{hidasi2016gru,LiCZY21,ZangLCLLYM22,AiWJG22,DongJ0LX21}. 
Though some \ac{SR} methods have achieved great success in many popular tasks, they meet the challenge while characterizing user preferences from the sparse data or cold-start scenarios. 
Therefore, \acf{CSR} is gaining increasing research attention that mitigates the above problem by leveraging the side information from other domains~\cite{ZhengSL022}.
The key idea of \ac{CSR} is to recommend the next-item to the overlapped user whose historical interactions can observe in multiple domains during the same period.

Early \ac{CSRs} incorporate a \ac{RNN}-based structure~\cite{ma2019pi,ren2021psjnet} to capture the sequential dependencies from the hybrid sequence but fail to model the associations among cross-domain entities. 
Then, the attention-based methods, e.g., Chen et al.~\cite{ChenGS21} and Li et al.~\cite{LiJQHT21}, adopt dual-attention structures that attentively transfer the users' sequential preference between domains but have difficulty excavating structural patterns inside the sequential transitions.
Recently proposed \acf{GCN}-based methods~\cite{ZhengSL022,guo2021dagcn,guo2022tida} for \ac{CSR} tasks bridge two domains by the \acf{CDS} graph and transfer the fine-grained domain knowledge by considering structural information.
However, the volume and computational complexity of such graph-based methods are enormous, resulting in the low running efficiency of models, which hinders their deployment on generic devices with limited memory (e.g., GPUs).

To realize the memory-efficient lightweight recommender system, recent studies~\cite{ChenYZHWW21,LiuZWLT20,MiaoLY22} predominantly focus on compressing the original collaborative filtering matrix to improve the recommendation efficiency.
However, the dependency between the origin representations might be disturbed in the process of embedding transformation, and the user's interest migration in a period also can not be sufficiently modeled.
Then, to explore the evolution and migration of users' preferences from a lightweight perspective, some researchers~\cite{LiCZY21,MeiZK22} shift their focus to transformer-based models.
Nevertheless, the self-attention mechanism within the Transformer~\cite{Vas2017attention} has quadratic complexity that brings a heavy parameter scale.
Similar problems also exist in \ac{CSRs}, even worse, because they need to share information across two domains, which leads to a doubling of the parameter size.
Hence, developing the lightweight \ac{CSRs} is an ongoing trend, but accompanied by two significant challenges: 
1) As mentioned earlier, the current \ac{CSRs} begin to use complex and in-depth models such as graph convolution networks~\cite{ZhengSL022,guo2021dagcn} to learn the primary preference for users, which brings a lot of burden for model training but contributes little to the node representation learning. 
2) The global relevance weighting strategy (e.g., the Transformer-based methods)~\cite{guo2022tida,LiCZY21,MeiZK22} may not be the optimal scheme to capture the sequential behavioral pattern, as it has enormous computational complexity and ignore the positional relations of the item from the dual-domain hybrid sequence.

In this work, we propose a novel \acf{LEA-GCN} to address the above challenges.
Concretely, we first extract the positional information of each item in the original hybrid sequence to better capture the inter-domain behavior evolution of overlapped users.
Then, we construct the \acf{CDS} graph and model the complicated inner-domain associations among users and items, such as the user-item interactions and the sequential orders of items in each domain.
After that, we take two steps to simplify the \ac{GCN} encoder: 
1) By removing the feature transformation matrices and the non-linear activation function (i.e., like the LightGCN~\cite{he2020lightGCN} does on NGCF~\cite{he2019ngcf}), we only keep the simple weighted sum aggregator in GCN to capture the collaborative filtering signals from both domains. 
2) Then, we adopt the \acf{SLAP}, which reduces the complexity of layer propagation and simultaneously avoids the interference caused by high-order connectivity in the \ac{CDS} graph.
To address the second challenge, we adopt a newly proposed technology, named \acf{EA}, which has attracted extensive research attention in the field of Computer Vision (CV)~\cite{FangYSWZ22,guo2022beyond}. 
\ac{EA} uses two external memory units to optimize the computational complexity of the traditional self-attention mechanism, which surprisingly matches our lightweight purpose.
However, the external storage units are independent of the input features, which leads to the deficiency in modeling items' collaborative filtering signal.
To avoid that, we devise a dual-channel \ac{EA}-based sequence encoder to learn the user-specific sequential pattern. 
It simultaneously calculates the correlation between items and the external memory units by a multi-head structure and the relation score of each item in the sequence via a \ac{MLP}.
The main contributions of this work can be summarized as follows:  
\begin{itemize}
\item After pointing out the defects of existing \ac{CSR} methods in parameter scale and training efficiency, we propose a lightweight GCN-based scheme, namely LEA-GCN, for the memory-efficient cross-domain sequential recommendation.
\item We improve the GCN by simplifying the network structure and using the single-layer aggregation protocols. Then, we devise a dual-channel external attention to model the user's sequential preference in a lightweight perspective.
\item Extensive experiments on two real-world datasets demonstrate that \ac{LEA-GCN} performs better and requires fewer parameters than several state-of-the-art baselines.
\end{itemize}

\section{Related Work \label{sec:relatedwork}}
\subsection{Sequential Recommendation}
\noindent As \acf{SRs} propose to model the user-item interaction sequence~\cite{LiCZY21}, it has been proven effective in capturing the evolution of user behavioral patterns~\cite{hidasi2016gru}. 
Existing studies on \ac{SR} can categorize into traditional methods and deep-learning-based methods. 
Early traditional \ac{SR} methods usually incorporate Markov chain assumption to capture high-order sequential patterns~\cite{HeKM17,ChengYLK13}. 
With the development of deep neural networks, researchers have applied the \ac{RNN}-based~\cite{quadrana2017hrnn,WuABSJ17}, \ac{GNN}-based~\cite{ZhengLLW20,ZangLCLLYM22,DongJ0LX21}, transformer-based~\cite{AiWJG22,KangM18,ChenLPWYLZWY22}, and self-supervised~\cite{QiuHYW22,xie2022contrastive} methods to the \ac{SR} task. 
These methods have the powerful capability of representation learning but have difficulty addressing the challenges caused by data sparsity or cold-start.
\subsection{Cross-domain Sequential Recommendation}
\noindent By treating the information from other domains as a supplement, the \acf{CSR} approaches can alleviate the data sparsity and the cold-start problems for \ac{SR}~\cite{MaRCRZLMR22}.
In early explorations, $\pi$-net~\cite{ma2019pi} and PSJNet~\cite{ren2021psjnet} are two \ac{RNN}-based solutions for \ac{CSR} that parallel share the information between domains and simultaneously learn the sequence representations for both of them.
Then, Zheng et al.~\cite{ZhengSL022} and Guo et al.~\cite{guo2021dagcn,guo2022tida} address \ac{CSR} from the graph-based perspective, which first builds the \ac{CDS} graph and attentively learns the user-specific representations in both local and global aspects. 
Another research direction is attention-based methods, such as Chen et al.~\cite{ChenGS21} and Li et al.~\cite{LiJQHT21}, which provide cross-domain recommendations by matching the user's sequential preference with candidate items
through a dual-attention learning mechanism. 
With the increasing depth of the neural networks and the complexity of the model structure, these \ac{CSR} methods are gradually becoming uncontrollable on the size of parameters or the memory overhead, exceeding the load of most conventional devices.
\subsection{Lightweight Recommendation}
\noindent To simplify the structure of the recommender system yet make it easier to be implemented on various devices, the concept of lightweight recommendation has attracted a lot of attention~\cite{MiaoLY22}. 
In traditional methods, recent studies have focused on lightening the structure of DeepFM~\cite{LiuGCJL21}, DNN~\cite{LiuZWLT20,LianWLLC020}, \ac{GCN}~\cite{MiaoLY22}, and transformer~\cite{MeiZK22} to improve the memory efficiency of recommenders. 
Li et al.~\cite{LiCZY21} are the first ones that introduce a lightweight solution for \acf{SR} via twin-attention networks to simultaneously address the challenges of lightening the parameter and discovering the temporal signals from all interacted items.
As \ac{CSR} usually requires auxiliary structures to bridge multiple domains, they often need more training time and a larger parameter scale, but the relevant lightweight solutions are mostly unexplored.
\section{Method \label{sec:method}}
\subsection{Preliminary \label{subsec:preliminary}}
\noindent \ac{CSR} task tends to recommend the next item for an overlapped user by modeling her/his historical interactions from the hybrid sequences~\cite{ZhengSL022}.
Suppose that $ U = \{U_1,U_2, \dots, U_k, \dots, U_p\}$ is the set of overlapped users whose historical behaviors are available in two domains, where $U_k \in\mathcal{U}$ $(1\leq k\leq p)$ denotes an independent user in $\mathcal{U}$. 
Let $S_H$ be the original hybrid sequence of an overlapped user, we further split the $S_H$ into $S_A=\{A_1, A_2, \dots, A_i, \dots, A_m\}$ and $ S_B=\{B_1, B_2, \dots, B_j, \dots, B_n\} $, which denote the interaction sequences in domain A and B respectively, where $A_i \in\mathcal{A}$ $(1\leq i\leq m)$ represents the items in domain A and $B_j \in \mathcal{B}$ $(1\leq j\leq n)$ represents the items in domain B.
Then we let $P_A=\{P_{A_1}, P_{A_2}, \dots, P_{A_i}, \dots, P_{A_m}\}$ and $ S_B=\{P_{B_1}, P_{B_2}, \dots, P_{B_j}, \dots, P_{B_n}\} $ be the positional information for the sequence $S_A$ and $S_B$ respectively, which accurately record the position of items from the original hybrid sequence $S_H$.

The probabilities of being recommended for all candidate items in both domains can be denoted as:
\begin{align}
& P(A_{i+1}|S_A, S_B)\sim f_A(S_A, S_B), \\
& P(B_{j+1}|S_B, S_A)\sim f_B(S_B, S_A),
\end{align}
where $P(A_{i+1}|S_A, S_B)$ is the probability of recommending $A_{i+1}$ as the next consumed item in domain A based on $S_A$ and $S_B$. And $f_A(S_A,S_B)$ denotes the learning function utilized to estimate the probability. And the similar definition for domain B can be denoted as $P(B_{j+1}|S_B, S_A)$ and $f_B(S_B, S_A)$.

\subsection{Overview \label{subsec:overview}}
\begin{figure*}[ht]
\label{fig:overview}
\centerline{\includegraphics[width=10cm]{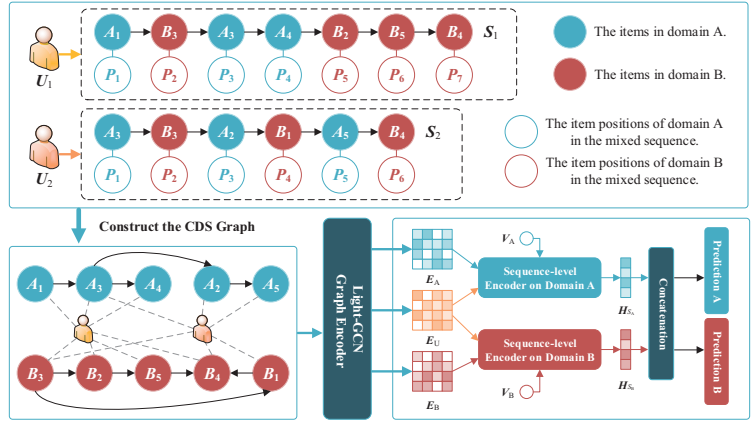}}
\caption{An overview of our proposed LEA-GCN, where $S_1$ and $S_2$ denote the hybrid sequence of two different overlapped users $U_1$ and $U_2$, respectively.}
\end{figure*}

\noindent The key idea of \ac{LEA-GCN} is to develop a lightweight graph-based solution for \ac{CSR} without affecting prediction accuracy. 
In \ac{LEA-GCN}, we optimize the structure of GCN by adopting the \acf{SLAP} and simplify the sequence encoder by taking advantage of the \acf{EA} mechanism.
As shown in Fig.~\ref{fig:overview}, we first record the position orders of all the items in the hybrid sequences to retain the information on users’ inter-domain behavioral patterns.
Secondly, by selecting items of each domain with the inner-domain sequential orders fixed from the hybrid sequences, we result in the subsequences $S_A$ and $S_B$ for domains A and B, respectively.
Then, we follow the same composition rules as DA-GCN~\cite{guo2021dagcn} to construct the \ac{CDS} graph by considering the sophisticated associations among users and items (i.e., inter-domain user-item interactive relations and inner-domain item-item order relations).
After constructing the \ac{CDS} graph, we adopt the Light-GCN~\cite{he2020lightGCN} graph encoder to linearly propagate embedding on the \ac{CDS} graph and adopt the \ac{SLAP} to learn the node representations and optimize the parameter scale synchronously.
Subsequently, to further capture the sequential patterns from users' interactions in a lower calculation complexity, we devise a dual-channel \acf{EA}-based sequence encoder. That calculates the correlation between all the items and considers the positional information (i.e., the $\bm{V}_A$ and $\bm{V}_B$) extracted from the hybrid sequence. 
Then, the resulting sequence-level representations can be denoted as $\bm{H}_{S_A}$ and $\bm{H}_{S_B}$ for both domains. We finally feed the concatenation of $\bm{H}_{S_A}$ and $\bm{H}_{S_B}$ to the prediction layer.
\subsection{Lightweight Node Representation Learning}
\subsubsection{Graph Construction.}
Inspired by Guo et al.~\cite{guo2022tida}, we construct the \ac{CDS} graph to link two domains by considering two types of associations: 1) user-item interactions between both domains; 2)item-item sequential transitions within both domains, where users and items in each domain are nodes and their associations are edges. However, such a composition method only retains the inner-domain sequential transferring characteristics, but discards the order dependency of items from the original hybrid sequences. To fix the above defects, we additionally record the positional information of items in the hybrid sequence before constructing the \ac{CDS} graph, so as to use them in the sequence-level representation learning.

Then, the \ac{CDS} graph $\bm{\mathcal{G}}\in \mathbb{R}^{(m+p+n)\times(m+p+n)}$ can be described in a matrix-form as:
\begin{align}
& \bm{\mathcal{G}}={
\left[ \begin{array}{cc}
    \bm{A} & \bm{0} \\
    \bm{0}     & \bm{B},
\end{array} 
\right ]}; 
& \bm{A}={
\left[ \begin{array}{cc}
    \bm{0} & \bm{R}_{AU} \\
    \bm{R}_{AU}^{T}     & \bm{0},
\end{array} 
\right ]}; \qquad \quad
& \bm{B}={
\left[ \begin{array}{cc}
    \bm{0} & \bm{R}_{BU} \\
    \bm{R}_{BU}^{T}     & \bm{0},
\end{array} 
\right ]},
\end{align}
where $\bm{A} \in \mathbb{R}^{(m+p)\times(m+p)}$ and $\bm{B} \in \mathbb{R}^{(p+n)\times(p+n)}$ respectively denote the compressed Laplace matrices of both domains, $\bm{R}_{AU} \in \mathbb{R}^{m\times p}$ and $\bm{R}_{BU} \in \mathbb{R}^{p\times n}$ represent the user-item interaction matrix of domain A and B, respectively, $\bm{R}_{AU}^T \in \mathbb{R}^{m\times p}$ and $\bm{R}_{BU}^T \in \mathbb{R}^{p\times n}$ are the transpose matrices.

\subsubsection{Single-Layer Aggregation Protocol (SLAP).}
\noindent It has been proven effective in early proposed state-of-the-art methods NGCF~\cite{he2019ngcf} and LightGCN~\cite{he2020lightGCN}, which adopt a \acf{MLAP} to propagate embeddings from nodes layer by layer. 
The core idea of the \ac{MLAP} is to use the high-order connectivity on the user-item bipartite graph to obtain the potential association between them, thereby enhancing the performance of node representation learning. 
The node representation learning on the l-th layer can be detailed as:
\begin{align}
& \bm{E}^{(l)}=H(\bm{E}^{(l-1)}, \mathcal{G}),
\end{align}
where $\mathcal{G}$ is the user-item bipartite graph and $\bm{E}^{(l)}$ denotes the node representations at the l-th layer, $\bm{E}^{(l-1)}$ is that of the previous layer. $H(\cdot)$ represents the function for neighbor aggregation.

However, the \ac{MLAP} may not suitable for the \ac{CDS} graph.
For example, when learning the item representation of domain A by considering the high-order connectivity of the items in domain B. It will also bring high-order domain-specific structural information from domain B, which might be the noise message for domain A, interfering with the interest expression of overlapped users. 
Hence, we develop a single-layer propagation rule (a.k.a., the \acf{SLAP}) to consider first-order neighbor aggregation on the \ac{CDS} graph as:
\begin{align}
& \bm{E}_U=\sum_{k\in U}(\sum_{i\in S_A}\frac{1}{\sqrt{\lvert N_U\rvert} \sqrt{\lvert N_A\rvert}}\bm{e_{i\rightarrow k}} + \sum_{j\in S_B}\frac{1}{\sqrt{\lvert N_U\rvert} \sqrt{\lvert N_B\rvert}}\bm{e_{j\rightarrow k}});\\
& \qquad \qquad \qquad \bm{E}_A=\sum_{i\in S_A}\sum_{k\in U}\frac{1}{\sqrt{\lvert N_U\rvert} \sqrt{\lvert N_A\rvert}}\bm{e_{k\rightarrow i}};\\
& \qquad \qquad \qquad \bm{E}_B=\sum_{j\in S_B}\sum_{k\in U}\frac{1}{\sqrt{\lvert N_U\rvert} \sqrt{\lvert N_B\rvert}}\bm{e_{k\rightarrow j}},
\end{align}
where $\bm{e_{i\rightarrow k}}$ and $\bm{e_{j\rightarrow k}}$ denote the passing message from the items to the overlapped user, $\bm{e_{k\rightarrow i}}$ and $\bm{e_{k\rightarrow j}}$ respectively denote the message transferred from users to items of domain A and domain B, $N_U$ is the set of overlapped users that interact with item $A_i$ or item $B_j$, $N_A$ and $N_B$ are the set of items that are interacted by user $u$. By adopting the \ac{SLAP}, the knowledge between domains could be transferred with less impact from the domain-specific information and simultaneously reduces the parameter scale. 

To support our view, we conduct a series of ablation experiments to investigate the impact of the layer depth. Due to the space limitation, we only report the experimental results of DOUBAN (i.e., a real-world dataset which will be detailed in Section~\ref{subsub:datasets}) on Table~\ref{tab:layers}.
\begin{table}
  \centering
  \caption{Performance ($\%$) comparison between LightGCN, DA-GCN and our proposed GCN-based solution at different layers.}
  \small
  \begin{tabular}{c|c|ccc|ccc}
    \toprule
    \multicolumn{2}{c|}{\textbf{Domain}}& \multicolumn{3}{c|}{\textbf{A}} &
    \multicolumn{3}{c}{\textbf{B}}\\
    \midrule
    \multicolumn{1}{c|}{\textbf{Layer Numbers}}& \multicolumn{1}{c|}{\textbf{Method}}&\multicolumn{1}{c}{RC10} & \multicolumn{1}{c}{MRR10} 
    &\multicolumn{1}{c|}{NDCG10}& \multicolumn{1}{c}{RC10} & \multicolumn{1}{c}{MRR10} 
    &\multicolumn{1}{c}{NDCG10}  
          \\
    \midrule
    \multicolumn{1}{c|}{\multirow{3}[1]{*}{\textbf{@ 1 Layer}}}
    & LightGCN &78.03&75.06&48.24
    &68.28&52.03&37.52\\
    &DA-GCN &83.55 &80.84 &51.53
    &71.91 &58.31 &40.67 \\
    &\textbf{LEA-GCN} &\textbf{83.83} & \textbf{81.22} & \textbf{52.06}
    &\textbf{76.14} & \textbf{66.15} & \textbf{45.01} \\
    \midrule
    \multicolumn{1}{c|}{\multirow{3}[1]{*}{\textbf{@ 2 Layers}}}
    & LightGCN &78.33&75.25&48.89
    &68.19&52.00&37.66\\
    & DA-GCN &83.12&80.55&50.92
    &71.42&57.45&40.16\\
    & LEA-GCN &83.53&81.08&51.35
    &74.63&63.67&42.91\\
    \midrule
    \multicolumn{1}{c|}{\multirow{3}[1]{*}{\textbf{@ 3 Layers}}}
    & LightGCN &78.52 &75.35 &48.93 &68.32 &52.11 &37.79  \\
    & DA-GCN &81.99&80.08&49.35
    &71.08&57.17&39.85\\
    & LEA-GCN &83.40&81.02&51.05
    &74.22&63.29&42.11\\
    \bottomrule
    \end{tabular}%
  \label{tab:layers}%
\end{table}%

As shown in Table~\ref{tab:layers}, we search the performance of \ac{LEA-GCN} at different layers compared with two state-of-the-art GCN-based methods (i.e., LightGCN~\cite{he2020lightGCN} and DA-GCN~\cite{guo2021dagcn}). To make the LightGCN comparable to other two \ac{CSR} methods, we simultaneously report its performance on both domains.
Then, we have the following observations: 1) Increasing the number of layers can improve the performance of LightGCN, demonstrating the effectiveness of considering the high-order connectivity while propagating embedding on the user-item graph within a single domain. 
2) DA-GCN and LEA-GCN with \ac{SLAP} perform better than them with \ac{MLAP}, which supports our hypothesis that simply aggregates message from the high-order connectivity on the \ac{CDS} graph may bring more noise to the node representation learning and proves the significance of adopting \ac{SLAP} for the \ac{CSR} task.

\subsection{External Attention (EA)-based Sequence Encoder}
\noindent Self-attention uses the combination of self values to refine the input sequence representations, which only considers the relation between items within a sequence but ignores implied relationships between items in different sequences~\cite{guo2022beyond}. And the high computational complexity of O($N^2$) presents another significant drawback to use self-attention. 
To accurately measure different items' contributions to the sequence yet with lower computational costs, we replace the self-attention-based algorithms~\cite{guo2022tida,MeiZK22} by \acf{EA} mechanism which have achieved great success on the task of image classification, object detection, and semantic segmentation~\cite{guo2022beyond}. 
\begin{figure*}[ht]
\label{fig:seqencoder}
\centerline{\includegraphics[width=10cm]{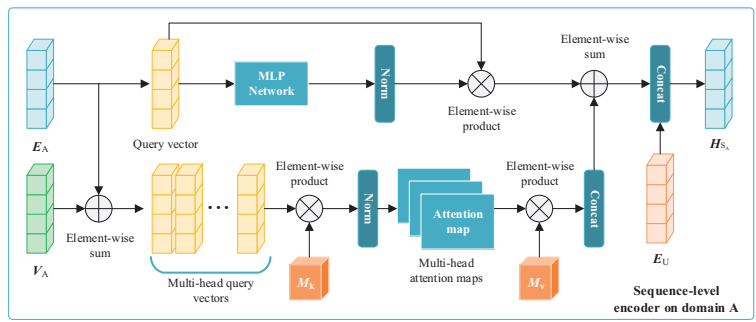}}
\caption{The workflow of our proposed dual-channel external attention-based sequence encoder (take domain A as an example), where $\bm{M}_k$ and $\bm{M}_v$ are two different external memory units.}
\end{figure*}

Different from the self-attention mechanism, \acf{EA} uses external storage units to reserve the global sharing weights. 
As shown in Fig.~\ref{fig:seqencoder}, We devise a dual-channels composite linear structure to implement the \ac{EA} (take domain A as an example). 
Specifically, in the first channel, we attach the positional information $\bm{V}_A$ from the hybrid sequence to the item representation $\bm{E}_A$. Then, we calculate the external attention between items and two external memory units (i.e., $\bm{M}_k$ and $\bm{M}_v$), which respectively act as the key and value matrices. The score calculations can be formulated as:
\begin{align}
  \bm{T}_{S_A}^{(1)}=Norm[(\bm{E}_A + \alpha\bm{V}_A)\bm{M}_k^T];\\
 \bm{H}_{S_A}^{(1)}=\bm{T}_{S_A}^{(1)}\bm{M}_v^T, \qquad\quad
\end{align}
where $\alpha$ is a hyper-parameter that controls the participation of the positional information, $\bm{T}_{S_A}^{(1)}$ is the external attention map, $\bm{H}_{S_A}^{(1)}$ represents the final output embedding of sequence $S_A$ from the first channel. For the i-th item $A_i$ in $S_A$, the calculation can be further detailed as:
\begin{align}
    \bm{a}_{i,j}^{(1)}=Norm[(e_{A_i} \oplus (\alpha\cdot v_{A_i}))m_j^k];\\
    \bm{H}_{S_A}^{(1)}=\sum_{i=1}^{\lvert S_A\rvert}\bm{a}_{i,j}^{(1)}m_i^v,\qquad
\end{align}
where $\bm{a}_{i,j}^{(1)}$ is the pair-wise affinity between i-th items $e_{A_i}$ and the j-th row $m_j^k$ of matrix $M_k$. $v_{A_i}$ denotes the positional information of $A_i$ and $m_i^v$ denotes the i-th row of the second memory unit.

Inspired by the Transformer~\cite{Vas2017attention}, we improve the capacity of \ac{EA} by adopting a multi-head manner~\cite{guo2022beyond} to better capture different relations between items as:
\begin{align}
    Z_h^{(1)}=ExternalAttention(\bm{E}_A, \bm{V}_A, \bm{M}_k, \bm{M}_v);\\
    \bm{H}_{S_A}^{(1)}=Concat(Z_h^{(1)}, \dots, Z_{\beta}^{(1)})\bm{W}_1,\qquad
\end{align}
where $\bm{W}_1$ is the linear transformation matrix to align the dimensions of input and output, $Z_h$ is the h-th head, and $\beta$ controls the number of heads. In experiments, we search $\beta$ in [1, 2, 4, 8, 16] and report the experimental results in Section~\ref{subsec:paramanalysis}.

Although external attention can well solve the problem of computational complexity, its final output representation is generated by $M_v$. As the external memory units are independent of the input features (i.e., the $E_A$), it will lead to the deficiency in modeling items’
collaborative filtering signals. 
To address above questions, in the second channel, an \ac{MLP} network with a smooth normalization layer (Inspired by Guo et al.~\cite{guo2022tida}) is used to extract the collaborative filtering signals of $E_A$. We measure the correlation between $e_i$ and $e_j$ as:
\begin{align}
    \bm{a}_{i,j}^{(2)}=Norm(f(e_{A_i}, e_{A_j}))=Norm(\bm{W}_3^T\bm{ReLU}(\bm{W}_2[e_{A_i}\oplus e_{A_j}]+\bm{b})),
\end{align}
where $f(\cdot)$ is a score function implemented by an \ac{MLP} network, $\bm{W}_2$ and $\bm{W}_3$ are two weight matrices, and $\bm{b}$ is the bias vector.
Then we can get the sequence representation $\bm{H}_{S_A}^{(2)}$ from the second channel:
\begin{align}
    \bm{H}_{S_A}^{(2)}=\sum_{i=1}^{\lvert S_A\rvert}\bm{a}_{i,j}^{(2)}e_{A_i}.
\end{align}

Hence, the resulting user-specific sequence-level representation $\bm{H}_{S_A}$ for domain A can be denoted as:
\begin{align}
    \bm{H}_{S_A} = Concat((\bm{H}_{S_A}^{(1)} + \bm{H}_{S_A}^{(2)}), \bm{E_U}).
\end{align}

\subsection{Prediction Layers}
\noindent After the sequence representation learning, \ac{LEA-GCN} gets the sequence embedding $\bm{H}_{S_A}$ and $\bm{H}_{S_B}$ for domain A and B, respectively. Then, for leveraging the information in both domains, we feed the concatenation of them to the prediction layer:
\begin{align}
P(A_{i+1}|S_A, S_B) = softmax(\bm{W}_A \cdot [\bm{H}_{S_A},\bm{H}_{S_B}]^\mathrm{T}+\bm{b}_A);\\
P(B_{j+1}|S_B, S_A) = softmax(\bm{W}_B \cdot [\bm{H}_{S_B},\bm{H}_{S_A}]^\mathrm{T}+\bm{b}_B),
\end{align}
where $\bm{W}_A$ and $\bm{W}_B$ are the weight matrix of all items in domain A and B, respectively; $\bm{b}_A$ and $\bm{b_A}$ are the bias term for both domains. Then, to avoid the seesaw phenomenons in $\pi$-net~\cite{ma2019pi} and DA-GCN~\cite{guo2021dagcn}, we adopt the cross-entropy loss and optimize them independently on both domains:
\begin{align}
\mathcal{L}_A = -\frac{1}{|\mathcal{S}|}\sum_{S_A, S_B \in \mathcal{S}}\sum_{A_i \in S_A}\text{log} P(A_{i+1}|S_A,S_B), \\
\mathcal{L}_B = -\frac{1}{|\mathcal{S}|}\sum_{S_B, S_A \in \mathcal{S}}\sum_{B_j \in S_B}\text{log}P(B_{j+1}|S_B,S_A),
\end{align}
where $\mathcal{S}$ denotes the training sequences in both domains.
\section{Experiment \label{sec:experiment}}
\noindent We conduct extensive experiments on two real-world datasets to validate the effectiveness of \ac{LEA-GCN}. In this section, we aim to answer the following Research Questions (RQ):
\begin{itemize}
    \item[\textbf{RQ1:~}]Does \ac{LEA-GCN} work on lightening the model's weights? How is the training efficiency of the \ac{LEA-GCN}?
    \item[\textbf{RQ2:~}]How does the \ac{LEA-GCN} perform compared with other state-of-the-art baselines? Does our lightweight strategy lead to the deterioration of recommendation performance?
    \item[\textbf{RQ3:~}]Is it helpful to consider the items’ positional relationship in the hybrid input sequences? Does it work by using external attention to learn sequence representation for both domains? 
    \item[\textbf{RQ4:~}]How do the hyper-parameters affect the performance of \ac{LEA-GCN}?
\end{itemize}
\subsection{Experimental setup \label{subsec:experimentalset}}

\subsubsection{Datasets and Evaluation Protocols.~\label{subsub:datasets}}
We evaluate \ac{LEA-GCN} on two real-world datasets (i.e., DOUBAN~\cite{ZhuangZYZAXHX20} and AMAZON~\cite{fu2019ccr}). DOUBAN contains historical interactions of overlapped users on domain A and domain B (i.e., douban movies and douban books), which are collected from the well-known Chinese social media platform Douban\footnote{http://www.douban.com/}~\cite{ZhuangZYZAXHX20}. AMAZON is a product review dataset collected by Fu et al.~\cite{fu2019ccr}. It contains overlapped users' review behaviors on two different amazon\footnote{http://jmcauley.ucsd.edu/data/amazon/} platforms, i.e., amazon-book (domain A) and amazon-movie (domain B). As shown in Table~\ref{tab:dataset_statistics}, We randomly choose 80$\%$ of all the hybrid sequences of both datasets as the training sets, and the rest 20$\%$ as the testing sets. Moreover, for the pretreatments on both datasets, we filter out the cold users with less than ten historical interactions and those cold items which only noticed less than five times~\cite{hidasi2016gru}.

For evaluation, we first treat the last two observed items in each hybrid sequence as the ground truth items for both domains. Secondly, we employ three frequently used metrics (i.e., RC@10, MRR@10, and NDCG@10)~\cite{guo2021dagcn,guo2022rlisn} to evaluate each instance on the testing sets and report their average values.

\begin{table}
    \centering
    \small
     \caption{Statistics of the datasets, where A and B represent different domains for both datasets.}
    \begin{tabular}{lcc|cc}
    \toprule
    \multicolumn{1}{c}{\textbf{Dataset}}&\multicolumn{2}{c|}{\textbf{DOUBAN}}&       \multicolumn{2}{c}{\textbf{AMAZON}} \\
    \cmidrule{1-5}
    \multicolumn{1}{c}{\textbf{Domain}}
    & \textbf{A} & \textbf{B}
    & \textbf{A} & \textbf{B} \\
    \midrule
    Items &14,636 &2,940 &126,526 &61,362\\
    Interactions &607,523 &360,798 &1,678,006 &978,226\\
    \midrule
    Users & \multicolumn{2}{c|}{6,582}
    & \multicolumn{2}{c}{9,204} \\
    Sequences (Train) & \multicolumn{2}{c|}{42,062}
    & \multicolumn{2}{c}{90,574} \\
    Sequences (Test) & \multicolumn{2}{c|}{10,431}
    & \multicolumn{2}{c}{14,463} \\
    \bottomrule
    \end{tabular}
    \label{tab:dataset_statistics}
\end{table}

\subsubsection{Baselines. \label{subsub:baselines}}
To validate the performance of \ac{LEA-GCN}, we compared our proposed method with the following baselines: 
1) Traditional recommendations: NCF~\cite{he2017ncf}, NGCF~\cite{he2019ngcf}, and LightGCN~\cite{he2020lightGCN}. We adapt the traditional methods with sequential inputs and report their experimental results in each domain.
2) Sequential recommendations: GRU4REC~\cite{hidasi2016gru} and HRNN~\cite{quadrana2017hrnn}. We report their performance in each domain.
3) Cross-domain Sequential recommendations: $\pi$-Net~\cite{ma2019pi}, PSJNet~\cite{ren2021psjnet}, DA-GCN~\cite{guo2021dagcn}, and TiDA-GCN~\cite{guo2022tida}.

\subsubsection{Implementation Details. \label{subsub:implementation}}
We implement \ac{LEA-GCN}\footnote{https://github.com/JinyuZ1996/LEA-GCN
} by TensorFlow and accelerate the model training by NVIDIA Tesla K80M GPU. For parameters, we employ Xavier~\cite{glorot2010xavier} for initialization and optimize them by Adam~\cite{kingma2014adam}. 
To train the model, we set the batch-size as 256, the dropout ratio as 0.1, and the learning rate as 0.002 for domain A and 0.004 for domain B, respectively. For LEA-GCN, we set the embedding size as 16 and the regularization ratio as 1e-7. The hyper-parameter $\alpha$ is searched in [0-1] with a step size of 0.1 to adjust the participation of the positional information and the number of attention head $\beta$ is explored within [1, 2, 4, 8, 16] to reach the best performance for \ac{LEA-GCN}. We detail the experimental results for $\alpha$ and $\beta$ in \ref{subsec:ablation}. Moreover, we uniformly set the embedding-size to 16 for all the reference baselines to make their results comparable. As for other hyper-parameters, we refer to the best settings of their papers and fine-tune them on both datasets.

\subsection{Parameter Scale $\&$ Training Efficiency (RQ1) \label{subsec:training}}
\begin{figure}
    \centering
    \centerline{\includegraphics[width=12cm]{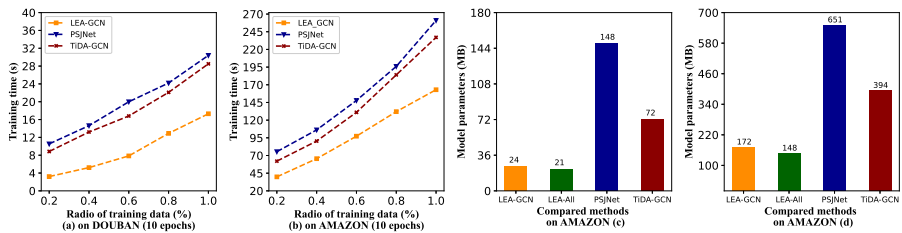}}
    \caption{Time consumption and the parameter scale of \ac{LEA-GCN} compared with PSJNet and TiDA-GCN.}
    \label{fig3:time_cost}
\end{figure}

\noindent In this section, we first conduct a series of experiments by changing the ratio of input data in [0.2 - 1.0] on DOUBAN and AMAZON to measure the time consumption of the model training.
Second, we analyze the performance of \ac{LEA-GCN} in lightweight modeling ability by measuring the scale of its parameters compared with two most competitive baselines (i.e., PSJNet and TiDA-GCN).
Then we have the following observations: 1) From Fig.~\ref{fig3:time_cost} (a) and (b), we notice that \ac{LEA-GCN} costs lower training time than TiDA-GCN and PSJNet, which demonstrates that \ac{LEA-GCN} has a better training efficiency and is scalable to the large-scale datasets.
2) From Fig.~\ref{fig3:time_cost} (c) and (d), we observe that \ac{LEA-GCN} and LEA-All need far fewer parameters than PSJNet and TiDA-GCN, providing a positive answer to RQ1. Note that, the LEA-All is a variant method that only keeps the GCN encoder in \ac{LEA-GCN}.

\subsection{Performance on Recommendation Accuracy (RQ2)\label{subsec:overallperform}}
\noindent Table~\ref{tab:resutls} shows the experimental results of \ac{LEA-GCN} compared with other state-of-the-art methods on both datasets. The observations are summarized as follows: 
1) \ac{LEA-GCN} achieves the best performance on both domains of AMAZON and outperforms other state-of-the-art baselines in most evaluation metrics on DOUBAN, demonstrating that our lightweight strategy in \ac{LEA-GCN} has little impact on the prediction accuracy, even improves the model's performance. 
2) The \ac{CSR} solutions outperform other state-of-the-art methods (i.e., traditional recommenders and sequential recommenders), demonstrating the significance of simultaneously modeling users’ sequential preference and cross-domain characteristics.
3) \ac{LEA-GCN} outperforms all the \ac{CSR} baselines, indicating the effectiveness of modeling user-specific preference from the \ac{CDS} graph by a lightweight structure, and displays the superiority of external attention in capturing users' sequential patterns. 

\begin{table*}
  \centering
  \normalsize
   \caption{Experiment results (\%) of compared methods on DOUBAN and AMAZON. Note that, the bold value denotes the best result in terms of the corresponding metric. Significant improvements are marked with$^\dag$ (paired samples t-test, $p<.05$).}
   \resizebox{\linewidth}{!}{ 
    \begin{tabular}{lcccccc|cccccc}
    \toprule
    \multicolumn{1}{c}{\textbf{Dataset}} & \multicolumn{6}{c|}{\textbf{DOUBAN}} &
    \multicolumn{6}{c}{\textbf{AMAZON}} \\
    \cmidrule{1-13}
    \multicolumn{1}{c}{\textbf{Domain}}& \multicolumn{3}{c}{\textbf{A}} & \multicolumn{3}{c|}{\textbf{B}} & \multicolumn{3}{c}{\textbf{A}} & \multicolumn{3}{c}{\textbf{B}} \\
    \cmidrule{1-13}
    \multicolumn{1}{c}{\textbf{Metric (@10)}}& \multicolumn{1}{c}{\textbf{RC}} & \multicolumn{1}{c}{\textbf{MRR}} & \multicolumn{1}{c}{\textbf{NDCG}}
          & \multicolumn{1}{c}{\textbf{RC}} & \multicolumn{1}{c}{\textbf{MRR}}
          & \multicolumn{1}{c|}{\textbf{NDCG}}&
          \multicolumn{1}{c}{\textbf{RC}} & \multicolumn{1}{c}{\textbf{MRR}}& \multicolumn{1}{c}{\textbf{NDCG}}
          & \multicolumn{1}{c}{\textbf{RC}} & \multicolumn{1}{c}{\textbf{MRR}} & \multicolumn{1}{c}{\textbf{NDCG}}
          \\
    \midrule
    NCF &69.75 &58.05 &40.26 &35.24 &23.29 &18.28 
    &15.59 &11.30 &6.12 &17.38 &13.20 &5.29\\
    NGCF &79.21&77.82&49.12&67.37&54.41&37.26
    &21.52&18.74&12.08&26.55&25.37&17.45\\
    LightGCN &78.52 &75.35 &48.93 &68.32 &52.11 &37.79 
    &21.67 &17.41 &10.13 &26.49 &24.23 &15.09\\
    \midrule
    GRU4REC &80.13& 75.41 &47.81 &66.66 &54.10 &38.03
    &21.51 &17.11 &9.66 &24.51 &22.13 &12.94\\
    HRNN &81.25 &77.90 &48.88 &68.32 &54.99 &38.93 
    &21.92 &17.30 &9.87 &25.10 &22.48 &13.44\\
    \midrule
    $\pi$-net &83.22&80.71&51.22&69.54&55.72&39.18
    &24.33 &20.52 &11.80 &27.66 &25.03 &16.20\\
    PSJNet &83.54&80.96&51.72&71.59&58.36&40.71
    &25.03 &21.09 &13.54 &31.24 &28.35 &18.93\\
    DA-GCN &83.55 &80.84 &51.53 &71.91 &58.31 &40.67
    &24.62 &20.91 &13.18 &31.12 &28.21 &18.85\\
    TiDA-GCN &83.68 &\textbf{81.27} &52.02 &72.56 &60.27 &41.38
    &25.05 &21.23 &14.68&32.84 &29.65 &19.12\\
    \midrule
    \textbf{LEA-GCN} &$\textbf{83.83}^\dag$ &81.22 &\textbf{52.06}
    &$\textbf{76.14}^\dag$ &$\textbf{66.15}^\dag$ &$\textbf{45.01}^\dag$
    &$\textbf{25.47}^\dag$ &$\textbf{21.57}^\dag$ &$\textbf{14.93}^\dag$
    &$\textbf{33.97}^\dag$ &$\textbf{30.65}^\dag$ &$\textbf{20.46}^\dag$\\
    \bottomrule
    \end{tabular}%
    }
  \label{tab:resutls}%
\end{table*}%

\subsection{Ablation Study (RQ3) \label{subsec:ablation}}
\noindent In this section, we conduct a series of ablation studies on DOUBAN and AMAZON to explore the impact of different components on \ac{LEA-GCN}. Due to space limitations, we only report the results on DOUBAN.
\begin{table}
  \centering
  \caption{The experimental results ($\%$) of ablation studies on the DOUBAN dataset.}
  \small
  \begin{tabular}{lccc|ccc}
    \toprule
    \multicolumn{1}{c}{\textbf{Domain}}& \multicolumn{3}{c|}{\textbf{A}} &
    \multicolumn{3}{c}{\textbf{B}}\\
    \cmidrule{1-7}
    \multicolumn{1}{c}{\textbf{Metric (@10)}}& \multicolumn{1}{c}{RC} & \multicolumn{1}{c}{MRR} 
    &\multicolumn{1}{c|}{NDCG}& \multicolumn{1}{c}{RC} & \multicolumn{1}{c}{MRR} 
    &\multicolumn{1}{c}{NDCG}  
          \\
    \midrule
    LEA-Pos &83.67&81.16&52.00
    &75.10&64.35&44.38\\
    LEA-EA &82.62 &79.14 &50.43
    &72.82 &61.92 &42.41 \\
    LEA-All &82.12 &78.10 &49.64
    &71.28 &59.43 &41.28 \\
    \midrule
    \textbf{LEA-GCN} &\textbf{83.83} &\textbf{81.22} &\textbf{52.06}
    &\textbf{76.14} &\textbf{66.15} &\textbf{45.01} \\
    \bottomrule
    \end{tabular}%
  \label{tab:ablation}%
\end{table}%
As shown in Table~\ref{tab:ablation}, the LEA-Pos is a variant that disables the participation of the positional information from the hybrid sequence. LEA-EA is another variant model that removes the external attention-based sequence encoder. LEA-All is a variant that disables both the positional information and the \ac{EA}. The observations of Table~\ref{tab:ablation} are summarized as follows: 1) \ac{LEA-GCN} outperforms LEA-All and LEA-Pos, demonstrating the importance of the positional information of items from the original hybrid sequence for learning overlapped users' sequential characteristics. 2) \ac{LEA-GCN} performs better than LEA-EA, demonstrating the contribution of the external attention component for sequence-level representation learning.

\subsection{Hyper-parameters Analysis (RQ4) \label{subsec:paramanalysis}}
\begin{figure}
    \centering
    \centerline{\includegraphics[width=12cm]{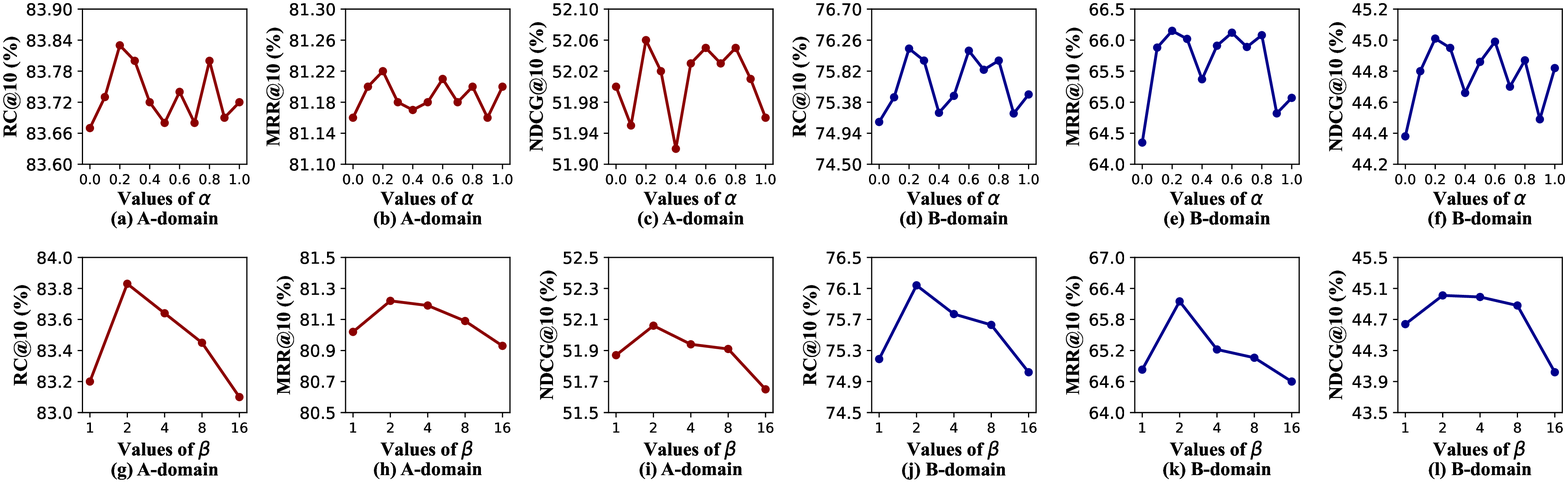}}
    \caption{Impact of hyper-parameters $\alpha$ and $\beta$ on DOUBAN.}
    \label{fig2:hyper_param}
\end{figure}

\noindent The hyper-parameter $\alpha$ controls the participation of the positional information attached to the item representations for both domains. Fig.~\ref{fig2:hyper_param} (a) to (f) show the performance of \ac{LEA-GCN} with different $\alpha \in $ [0 - 1]. The experimental results prove the significance of leveraging the positional information from the original hybrid sequences. However, it is not advisable to regard it as equally important as the sequence representation. 

The hyper-parameter $\beta$ controls the head number of the multi-head external attention component. The experimental results in Fig.~\ref{fig2:hyper_param} (g) to (l) demonstrate that only with an appropriate number of heads, does the \ac{EA} mechanism benefits the sequence representation learning process.
\section{Conclusions\label{sec:conclusion}}
\noindent In this work, we propose a lightweight GCN-based solution for \ac{CSR}, which simultaneously simplifies the structure of GCNs and optimizes the calculation complexity of the sequence encoder.
Specifically, we only keep the neighborhood aggregation to reduce the parameter scale of the GCN encoder and propose the \acf{SLAP} to propagate embedding on the \ac{CDS} graph.
Then, we devise a dual-channel \acf{EA}-based sequence encoder to calculate the correlation among all items via a lighter linear structure.
The experimental results on two real-world datasets demonstrate the superiority of our lightweight solution.


\bibliographystyle{splncs04}
\bibliography{0_Main}

\end{document}